\documentclass[aps,pra,amssymb,twocolumn,amsmath,superscriptaddress,showpacs,10pt]{revtex4}

\usepackage{graphicx}
\usepackage{dcolumn}
\usepackage{bm}
\usepackage{color}
\usepackage{epstopdf}

\begin{document}

\newcommand{\beq}{\begin{equation}}
\newcommand{\eeq}{\end{equation}}
\newcommand{\barr}{\begin{eqnarray}}
\newcommand{\earr}{\end{eqnarray}}

\newcommand{\REV}[1]{\textbf{\color{red}#1}}
\newcommand{\BLUE}[1]{\textbf{\color{blue}#1}}
\newcommand{\GREEN}[1]{\textbf{\color{green}#1}}
\newcommand{\RED}[1]{\textrm{\color{red}#1}}
\newcommand{\rev}[1]{{\color{blue}#1}}

\newcommand{\andy}[1]{ }
\newcommand{\bmsub}[1]{\mbox{\boldmath\scriptsize $#1$}}

\def\R{\mathbb{R}}

\def\bra#1{\langle #1 |}
\def\ket#1{| #1 \rangle}
\def\sinc{\mathop{\text{sinc}}\nolimits}
\def\cV{\mathcal{V}}
\def\cH{\mathcal{H}}
\def\cT{\mathcal{T}}
\def\cM{\mathcal{M}}
\def\cN{\mathcal{N}}
\def\CW{\mathcal{W}}
\def\e{\mathrm{e}}
\def\ii{\mathrm{i}}
\def\d{\mathrm{d}}
\renewcommand{\Re}{\mathop{\text{Re}}\nolimits}
\newcommand{\tr}{\mathop{\text{Tr}}\nolimits}

\title{Local Hamiltonians for Maximally Multipartite Entangled States}

\author{P. Facchi}
\affiliation{Dipartimento di Matematica, Universit\`a di Bari,
        I-70125  Bari, Italy}
\affiliation{INFN, Sezione di Bari, I-70126 Bari, Italy}
\affiliation{MECENAS, Universit\`a di Bari, I-70126 Bari, Italy}
\author{G. Florio}
\affiliation{Dipartimento di Fisica, Universit\`a di Bari,
        I-70126  Bari, Italy}
\affiliation{INFN, Sezione di Bari, I-70126 Bari, Italy}
\affiliation{MECENAS, Universit\`a di Bari, I-70126 Bari, Italy}
\author{S. Pascazio}
\affiliation{Dipartimento di Fisica, Universit\`a di Bari,
        I-70126  Bari, Italy}
\affiliation{INFN, Sezione di Bari, I-70126 Bari, Italy}
\affiliation{MECENAS, Universit\`a di Bari, I-70126 Bari, Italy}
\author{F. Pepe}
\affiliation{Dipartimento di Fisica, Universit\`a di Bari,
        I-70126  Bari, Italy}
\affiliation{INFN, Sezione di Bari, I-70126 Bari, Italy}

\date{\today}

\begin{abstract}
We study the conditions for obtaining maximally multipartite
entangled states (MMES) as non-degenerate eigenstates of
Hamiltonians that involve only short-range interactions. We
investigate small-size systems (with a number of qubits ranging
from $3$ to $5$) and show some example Hamiltonians with MMES as
eigenstates.
\end{abstract}


\maketitle

\section{Introduction}
\label{sec:intro}

The elusive features of multipartite entanglement are attracting
increasing attention lately. While in the bipartite case different
mathematical definitions are physically equivalent
\cite{entanglement,entanglementrev,h4}, a unique characterization
of multipartite quantum correlations does not exist, and
interesting alternative proposals are possible
\cite{multipart1,multipart2,multipart3,multipart4,Bergou}, that
highlight different features of this inherently quantum
phenomenon, including links with complexity and frustration \cite
{NJP}. The interest in multipartite entanglement is motivated by
possible applications in quantum enhanced tasks, but also by
genuine foundational aspects.

We proposed that the multipartite entanglement of a system of
qubits can be characterized in terms of the distribution function
of bipartite entanglement (e.g., purity) over all possible
bipartitions of the qubits \cite{MMES}. This led us to formulate
the notion of ``maximally multipartite entangled states" (MMES),
as those states for which average purity (over all balanced
bipartitions) is minimal. This notion can be extended to
continuous variable systems \cite{Adesso,FFLMP} and unearths novel
applications, such as quantum teamwork \cite{Adesso} and
controlled qubit teleportation \cite{zhasong}.

By their very definition, MMES exhibit very strong and
\emph{distributed} non-local correlations. This naturally leads to
the following question: can MMES be obtained by making use of
Hamiltonians that only involve \emph{local} interactions? In the
context of spin systems (that mostly concerns us here), ``local"
means both few-body and nearest-neighbors. For example, it would
be remarkable if one could find Hamiltonians containing up to
$2$-body interaction terms, whose ground state is a MMES. If this
were impossible, one could soften the requirement and ask whether
can one find Hamiltonians containing up to $2$-body interaction
terms, whose eigenstate is a MMES. This is the problem we intend
to tackle in the present paper. A similar problem was
analyzed in the context of GHZ states \cite{buzek,loss,loss1}.

This article is organized as follows. In Section
\ref{sec:potential} we review the notion of maximally multipartite
entangled state. In Section \ref{sec:frame} we sketch out our
general strategy for the search of MMES as eigenstates of
Hamiltonians with short range interactions. In Section
\ref{sec:example} we explicitly construct Hamiltonians having a
MMES as an eigenstate for a number of qubits ranging from $3$ to
$5$. This is of interest in few-qubit applications. We conclude
with a discussion on possible perspectives and applications.

\section{Maximally multipartite entangled states}
\label{sec:potential}

Let a system of $n$ qubits be in a pure state $|\psi\rangle \in
\mathcal{H}$, which is the only case we will consider henceforth.
We consider a partition $(A, \bar A)$ of the system $S=\{1,2,\dots
n\}$, made up of $n_A$ and $n_{\bar A}$ qubit, respectively, with
$n_A+n_{\bar A}=n$ and $n_A<n_{\bar A}$ with no loss of
generality. Purity reads
\begin{equation}\label{eq:piAdef}
\pi_A = \tr (\rho_A^2),
\end{equation}
where
\begin{equation}
\rho_A = \tr_{\mathcal{H}_{\bar{A}}} (|\psi\rangle\langle\psi|)
\end{equation}
is the reduced density matrix of party $A$.
Purity ranges between
\begin{equation}
\label{eq:piArange} \frac{1}{2^{n_A}} \leq \pi_A \leq 1,
\end{equation}
where the upper bound $1$ is reached by unentangled, factorized
states with respect to the bipartition ($A,\bar{A}$). On the other
hand, the lower bound is obtained for maximally bipartite
entangled states, whose reduced  density matrix is completely
mixed:
\begin{equation}
\rho_A= \frac{1}{2^{n_A}}  \openone_A ,
\end{equation}
where $\openone_{A}$ is the identity operator on the Hilbert space
of subsystem $A$.

The extension of this treatment to the multipartite scenario is
based on the average purity (``potential of multipartite
entanglement") \cite{scott,MMES}
\begin{equation}
\label{eq:piave} \pi_{\mathrm{ME}}(\ket{\psi}) =
\frac{1}{C^n_{n_A}} \sum_{|A|=n_A}\pi_{A},
\end{equation}
where $C^n_{n_A}$ is the binomial coefficient, $|A|$ is the
cardinality of the set $A$ and the sum is over balanced
bipartitions $n_A=[n/2]$, $[\cdot]$ denoting the integer part. The
quantity $\pi_{\mathrm{ME}}$ in Eq.\ (\ref{eq:piave}) measures the
average bipartite entanglement over all possible balanced
bipartitions and inherits the bounds (\ref{eq:piArange}) (with
$n_A=[n/2]$)
\begin{equation}
\label{eq:pmeconstraint}
\frac{1}{2^{[n/2]}} \le \pi_{\mathrm{ME}}(\ket{\psi})\le 1.
\end{equation}
A \emph{maximally multipartite entangled state} (MMES)
\cite{MMES} $\ket{\varphi}$ is a minimizer of
$\pi_{\mathrm{ME}}$,
\begin{eqnarray}
\label{eq:minimizer}
& & \pi_{\mathrm{ME}}(\ket{\varphi})= \pi_0^{(n)},  \\
& & \pi_0^{(n)} = \min \{\pi_{\mathrm{ME}}(\ket{\psi})\; | \;
\ket{\psi}\in \mathcal{H}_S, \bra{\psi}\psi\rangle=1\}.
\nonumber
\end{eqnarray}
Given a quantum system whose (pure) state is a MMES, the density
matrix of each one of its subsystems $A\subset S$ is as mixed as
possible (given the constraint that the total system is in a pure
state), so that the information contained in a MMES is as
distributed as possible. The average purity (\ref{eq:piave}) is
related to the average linear entropy \cite{scott} and extends
ideas put forward in \cite{multipart4,parthasarathy}. This
quantity has also been used to discuss generalized global
entanglement in one-dimensional critical systems
\cite{rigolin1,rigolin2}.

We can also define \emph{perfect}  MMES, obtained when the lower
bound (\ref{eq:pmeconstraint}) is saturated
\begin{equation}
\label{eq:perfectmmes} \pi_0^{(n)} =\frac{1}{2^{[n/2]}} .
\end{equation}
We notice that a necessary and sufficient condition for a state to
be a perfect MMES is to be maximally entangled with respect to
balanced bipartitions. On the other hand, this requirement can be
too strong; it can be shown that perfect MMES do not exist for
$n>8$ \cite{scott}.

\section{General Strategy}\label{sec:frame}
The problem of finding a Hamiltonian involving local (two-body and
nearest-neighbor) interactions and on-site external magnetic
fields, one of whose eigenstates is a MMES is non-trivial. As a
matter of fact, as we explained in the Introduction,  MMES exhibit
strongly non-local correlations, which, in principle, could be
impossible to obtain by using only local terms. On the other hand,
it is trivial to find Hamiltonians involving $n$-body interaction
terms, whose ground state is an $n$-qubit MMES: consider $n$
qubits on a circle and the Hamiltonian
\begin{equation}
\label{ham}
H(\epsilon,\mathcal{K})=\epsilon\mathcal{P}+H^{\perp}(\mathcal{K}),
\end{equation}
where $\mathcal{P}=\ket\varphi\bra\varphi$ is the projection on
the MMES $\ket\varphi$ and $H^{\perp}$ is a Hermitian operator
depending on the set of parameters $\mathcal{K}$ and satisfying
\beq \mathcal{P} H^{\perp}\mathcal{P}=0. \eeq If $H^{\perp}=0$ and
$\epsilon<0$, the MMES $\ket\varphi$ is by construction the
non-degenerate ground state for $H$.

This simple observation enables us to define our problem more
precisely. The Hamiltonian (\ref{ham}) can be separated into a
local part, in the sense defined above, and a nonlocal part, that
contains all other interaction terms:
\begin{equation}
H(\epsilon,\mathcal{K})=H_{\mathrm{loc}}(\epsilon,\mathcal{K})+H_{\mathrm{nonloc}}(\epsilon,\mathcal{K}).
\end{equation}
Our desideratum is to find a set of parameters
$(\bar{\epsilon},\bar{\mathcal{K}})$ such that
$H_{\mathrm{nonloc}}(\bar{\epsilon},\bar{\mathcal{K}})=0$, so that
the MMES $\ket\varphi$ is a non-degenerate eigenstate (possibly
the ground state). Clearly, this requirement might be impossible
to satisfy. In the following we will consider some explicit
examples for systems of 3, 4, and 5 qubits.

\section{Encoding a MMES into the eigenstate of a Hamiltonian}\label{sec:example}

\subsection{Three qubits}
For a system of three qubits, MMES are equivalent by local
unitaries to the GHZ states:
\begin{eqnarray}\label{GHZ3basis1}
|G_1^{\pm}\rangle&=&\frac{1}{\sqrt{2}}\left(|000\rangle\pm|111\rangle\right),
\\ |G_2^{\pm}\rangle&=&\frac{1}{\sqrt{2}}\left(|001\rangle\pm|110\rangle\right),
\\|G_3^{\pm}\rangle&=&\frac{1}{\sqrt{2}}\left(|010\rangle\pm|101\rangle\right),
\\  |G_4^{\pm}\rangle&=&\frac{1}{\sqrt{2}}\left(|011\rangle\pm|110\rangle\right),
\end{eqnarray}
where we have used the conventions $\sigma^z\ket 0=\ket 0$ and
$\sigma^z\ket 1=-\ket 1$, $\sigma^z$ being the third Pauli matrix.
Since these states form a basis of the Hilbert space of the
system, the most generic Hamiltonian can be written as
\begin{equation}\label{Hgen}
H=\sum_{i=1}^{4} \left(\epsilon_i^+ \mathcal{P}_i^+ + \epsilon_i^-
\mathcal{P}_i^- \right)+H_\mathrm{M},
\end{equation}
where
\begin{equation}
\mathcal{P}_i^{\pm}=|G_i^{\pm}\rangle\langle G_i^{\pm}|\quad\mbox{with}\;i=1, 2, 3, 4
\end{equation}
are the projections on the GHZ basis states and $H_\mathrm{M}$ is
a Hermitian operator containing terms of the form
\beq
|G_i^{\pm}\rangle\langle G_j^{\pm}|+\textrm{H.c.} \quad\mbox{with}\;i\ne j.
\eeq
We notice that each projection can be
decomposed in two terms
\begin{equation}\label{pqc}
\mathcal{P}_i^{\pm}=\mathcal{Q}_i\pm \mathcal{C}_i ,
\end{equation}
where $\mathcal{Q}_i$ contains products of two Pauli
matrices, while $\mathcal{C}_i$ includes
cubic terms. For example, $\mathcal{P}_1^{+}$ can be decomposed in the following operators:
\begin{eqnarray}
\mathcal{Q}_1&=&\frac{1}{2}\left(|000\rangle\langle
000|+|111\rangle\langle
111|\right) , \nonumber\\
&=&\frac{1}{8}\left(\openone+\sigma_1^z\sigma_2^z+\sigma_2^z\sigma_3^z+\sigma_1^z\sigma_3^z\right)\\
\mathcal{C}_1&=&\frac{1}{2}\left(|000\rangle\langle
111|+|111\rangle\langle 000|\right)\nonumber\\
&=&\frac{1}{8}\left(\sigma_1^x\sigma_2^x\sigma_3^x-\sigma_1^x\sigma_2^y\sigma_3^y-\sigma_1^y\sigma_2^x\sigma_3^y-\sigma_1^y\sigma_2^y\sigma_3^x\right)\nonumber\\
\end{eqnarray}
The decomposition in Eq.\ (\ref{pqc}) enables us to rewrite the
Hamiltonian (\ref{Hgen}):
\begin{equation}
H=\sum_{i=1}^{4} \left(\epsilon_i^++\epsilon_i^-\right)
\mathcal{Q}_i +\sum_{i=1}^{4}
\left(\epsilon_i^+-\epsilon_i^-\right) \mathcal{C}_i+H_\mathrm{M}.
\end{equation}
We notice that cubic terms in the operators $\mathcal{C}_i$ are
absent in $H_\mathrm{M}$, and the $\mathcal{C}_i$'s are orthogonal
to each other
\begin{equation}
\mathcal{C}_i\mathcal{C}_j=0 \qquad \forall i\neq j.
\end{equation}
Thus, the only way to cancel such cubic terms is to impose
\beq
\epsilon_i^+=\epsilon_i^- \quad \forall i.
\eeq
This equality immediately implies that
\begin{equation}\label{nocub}
\langle G_i^+|H_2|G_i^+\rangle=\langle G_i^-|H_2|G_i^-\rangle ,
\quad \forall i ,
\end{equation}
where
\begin{equation}
H_2=\sum_{i,j,\alpha,\beta} C_{ij}^{\alpha\beta} \sigma_i^{\alpha}
\sigma_j^{\beta}
\end{equation}
denotes a \emph{local} Hamiltonian (cubic couplings are absent).
As a consequence, the state $\ket{G_1^+}$ (and any equivalent
state by local unitaries) can never be the non-degenerate ground
state of $H_2$. Indeed, let us suppose that $\ket{G_1^+}$ is the
ground state of $H_2$, whose spectrum ranges from $E_0$ to
$E_{\mathrm{max}}$. If $|G_1^-\rangle$ is also an eigenstate, then
$E_0$ is degenerate, since condition (\ref{nocub}) must hold; if
$|G_1^-\rangle$ is not an eigenstate, then we must have
\begin{equation}
E_0<\langle G_1^-|H_2|G_1^-\rangle<E_{\mathrm{max}}
\end{equation}
and thus, as a consequence of Eq.\ (\ref{nocub}), $\ket{G_1^+}$
cannot be the ground state.

This result shows that it is impossible for a three-qubits MMES to
be the non degenerate ground state of a local Hamiltonian. On the
other hand, we now try to understand whether there exists a
condition such that $\ket{G_1^+}$ is a non-degenerate (excited)
eigenstate. The most general two-body (local) Hamiltonian is
\begin{eqnarray}\label{H2}
H_2&=& \sum_{i< j}
(J_{ij}^x\sigma_i^x\sigma_j^x+J_{ij}^y\sigma_i^y\sigma_j^y+J_{ij}^z\sigma_i^z\sigma_j^z)\nonumber\\
&+& \sum_{i\neq j}(K_{ij}\sigma_i^x\sigma_j^y+X_{ij}\sigma_i^x\sigma_j^z+Y_{ij}\sigma_i^y\sigma_j^z)\nonumber\\
&+&\sum_{i} (h_i^x\sigma_i^x+h_i^y\sigma_i^y+h_i^z\sigma_i^z).
\end{eqnarray}
By explicit calculation, it is possible to see that $\ket{G_1^+}$
is an eigenstate of $H_2$ if and only if the parameters satisfy
the following conditions:
\begin{eqnarray}\label{cond3}
&&\sum_{i=1}^3 h_i^z=0 \\
&&\sum_{j\neq i}X_{ij}=0 \quad \mathrm{for}\quad i=1,2,3, \\
&&\sum_{j\neq i}Y_{ij}=0 \quad \mathrm{for}\quad i=1,2,3, \\
&&h_1^x=J_{23}^y-J_{23}^x\,, \quad h_2^x=J_{13}^y-J_{13}^x, \\
&&h_3^x=J_{12}^y-J_{12}^x\,,\quad h_1^y=K_{23}+K_{32}, \\
&&h_2^y=K_{13}+K_{13}\,, \quad h_3^y=K_{12}+K_{21}.\label{cond3final}
\end{eqnarray}
Thus, the Hamiltonian $H_2$ has 23 free parameters.

It is important to notice that the requirement that $\ket{G_1^+}$
be an eigenstate does not have any influence on the coupling
between two qubits along the $z$-axis; actually, terms of the form
$\sigma_i^z\sigma_j^z$ are the only ones that leave the GHZ state
invariant.

\begin{figure}
\includegraphics[width=0.45\textwidth]{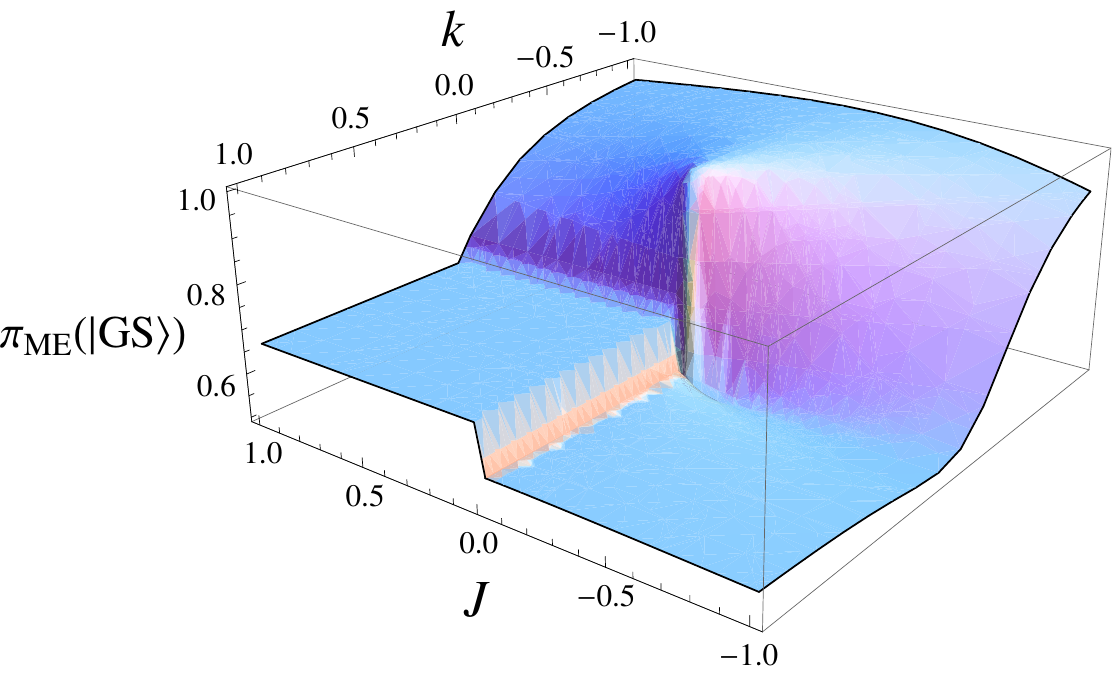}
\caption{(Color online) Potential of multipartite entanglement of
the ground state $\pi_{\mathrm{ME}}(|\textrm{GS}\rangle)$ for the
Hamiltonian (\ref{Hjk3}). The plateau in the region where $J$ and
$k$ are both positive corresponds to the constant value
$\pi_{\mathrm{ME}}=2/3$ when the ground state energy is $-(J+2k)$.
Elsewhere, the ground state corresponds to the eigenvalue
$J+2k-2\sqrt{J^2-2Jk+4k^2}$. }\label{gsent}
\end{figure}

To understand whether $\ket{G_1^+}$ is degenerate is not
straightforward. The task can be simplified by reducing the number
of free parameters satisfying Eqs.\
(\ref{cond3})-(\ref{cond3final}). We consider a model in which the
qubits are coupled along the $x$ and $z$ axes in a uniform
external field acting along $x$:
\begin{equation}\label{Hjk3}
H_{Jk}=J\sum_{i=1}^3 \sigma_i^z\sigma_{i+1}^z+ k\sum_{i=1}^3
\left(\sigma_i^x\sigma_{i+1}^x-\sigma_i^x\right),
\end{equation}
with periodic boundary conditions. The state
$\ket{G_1^+}$ corresponds to the eigenvalue $3J$, which is
non-degenerate if and only if
\begin{equation}
J\neq 0,\qquad k\neq 0,\qquad J\neq -\frac{k}{2}.
\end{equation}
As has been argued, the GHZ cannot be the ground state of
$H_{Jk}$: it corresponds to the first non-degenerate excited state
if the system is completely ferromagnetic ($J<0$ and $k<0$) or if
$k>0$ and $J<-k/2$. In the latter case, the ground state
$|\mathrm{GS}\rangle$ is not a MMES but has a large value of
entanglement, with $\pi_{\mathrm{ME}}\lesssim 0.556$ [Recall that
in Eq.(\ref{eq:perfectmmes}) $\pi_0^{(3)}=1/2$]. Thus, we have
found a range of values for the parameters of the local
Hamiltonian such that the two lowest energy states contain a large
amount of multipartite entanglement. Figure \ref{gsent} shows the
dependence on the parameters $J$ and $k$ of the potential of
multipartite entanglement $\pi_{\mathrm{ME}}$ for the ground state
of the Hamiltonian (\ref{Hjk3}).

\subsection{Four qubits}

\begin{figure}
\includegraphics[width=0.4\textwidth]{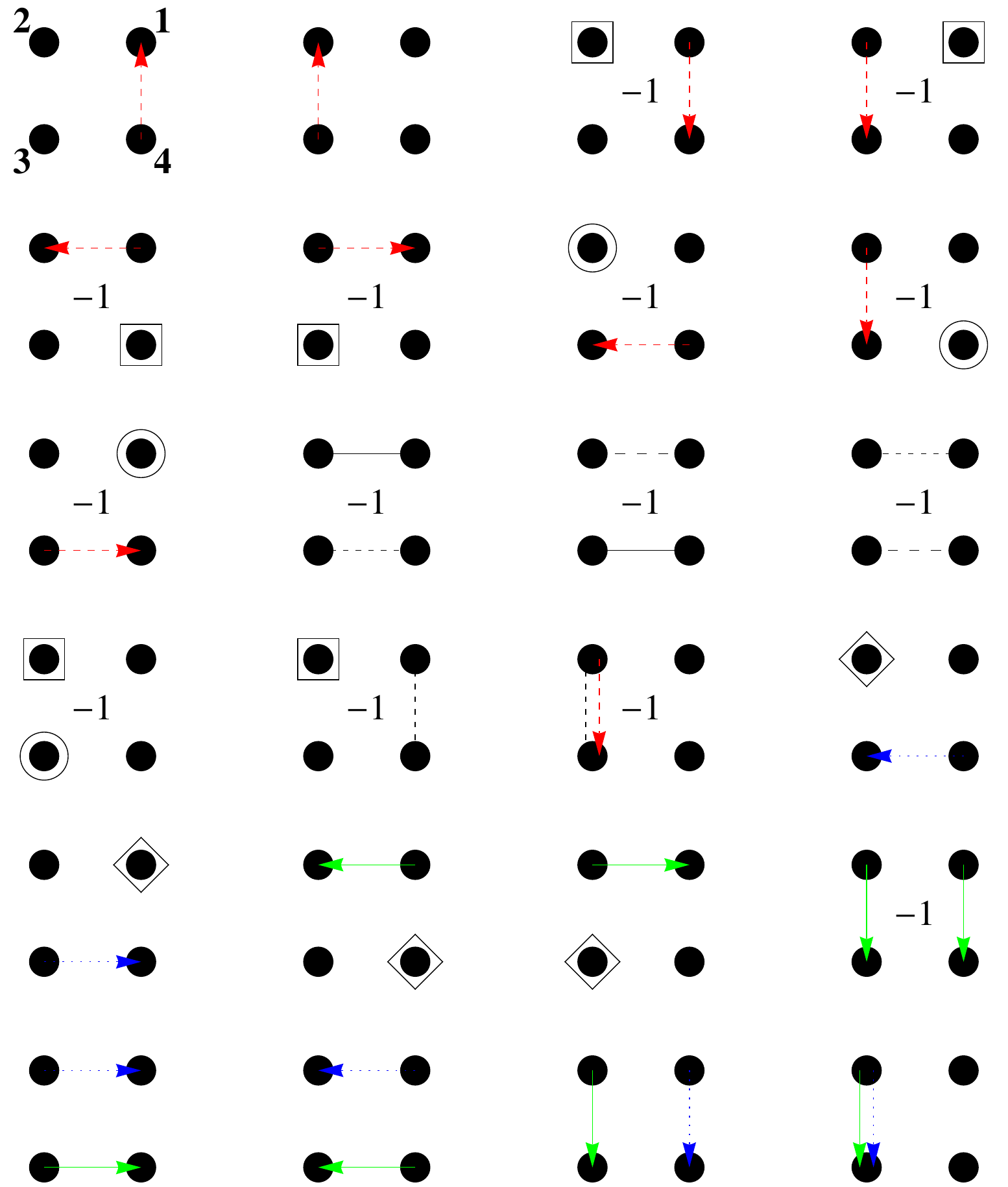}
\caption{(Color online) Graphic representation of all the terms of
a Hamiltonian having $|M_4^1\rangle$ as an eigenstate. The most
general Hamiltonian is a linear combination with arbitrary
coupling constants, of these terms.  In this representation, solid
lines correspond to couplings between the $x$-components of the
qubits, dotted lines between the $y$-components and dashed lines
between the $z$-components. Dashed (red) arrows from qubit $i$ to
qubit $j$ mean $\sigma_i^x\sigma_j^z$, dotted (blue) arrows
$\sigma_i^x\sigma_j^y$, full (green) arrows
$\sigma_i^y\sigma_j^z$. A circle around a qubit means an
interaction with an external field directed along $x$, a diamond
along $y$, a square along $z$. The number ``$-1$" inside a ring of
4 qubits means that the two contributions in the term must have
opposite coupling constants. The Hamiltonian (\ref{Hjk4}) is made
up of the the first six terms.}\label{hamilt4}
\end{figure}

In the case of four qubits it is known that perfect MMESs do not
exist \cite{MMES,scott,higuchi1,brown,higuchi2}. Numerical and
analytical analyses show that the minimum of the potential of
multipartite entanglement is $E_0^{(4)}=1/3>1/4$. In this
section we will search for local Hamiltonians having a
non-degenerate eigenstate (possibly the ground state),
corresponding to the uniform real state \cite{multent}
\begin{equation}\label{mmes4}
|M_4^1\rangle=\frac{1}{4}\sum_{i=0}^{15}\zeta^{(4)}_k|k\rangle
\end{equation}
determined by the coefficients
\begin{equation}\label{mmes4ph}
\zeta^{(4)}=\{1, 1, 1, 1, 1, 1, -1, -1, 1, -1, 1, -1, -1, 1, 1,
-1\}.
\end{equation}
This case can be treated in the same way as the three-qubit
system: we can construct a basis formed by 16 MMES, which will be
labelled as $|M_4^i\rangle$, all equivalent by local unitaries.
The generic Hamiltonian acting on the Hilbert space of the system
can be written as
\begin{equation}
H=\sum_{i=1}^{16} \epsilon_i \mathcal{M}_4^i +H_\mathrm{M}
\end{equation}
where $\mathcal{M}_4^i$ is the projection on the basis state $|M_4^i\rangle$. In
particular we have
\begin{eqnarray}\label{projmmes4}
\mathcal{M}_4^1&=&\frac{1}{16}  \bigl(
\mathbb{I}+\sigma_1^z\sigma_4^x+\sigma_2^z\sigma_3^x+\sigma_1^x\sigma_2^z\sigma_4^z+\sigma_1^x\sigma_3^x\sigma_4^z\nonumber\\
&+&\sigma_1^y\sigma_2^z\sigma_4^y
+\sigma_1^y\sigma_3^x\sigma_4^y
+\sigma_2^x\sigma_1^z\sigma_3^z+\sigma_2^x\sigma_3^z\sigma_4^x \nonumber\\
&+&\sigma_2^y\sigma_3^y\sigma_1^z
+\sigma_2^y\sigma_3^y\sigma_4^x+\sigma_1^x\sigma_2^x\sigma_3^y\sigma_4^y
 -\sigma_1^x\sigma_2^y\sigma_3^z\sigma_4^y\nonumber\\
 &+&\sigma_1^y\sigma_2^y\sigma_3^z\sigma_4^z
-\sigma_1^y\sigma_2^x\sigma_3^y\sigma_4^z+\sigma_1^z\sigma_2^z\sigma_3^x\sigma_4^z
\bigr).
\end{eqnarray}
By definition, $\epsilon_i$ are the expectation values on the
basis states and $H_\mathrm{M}$ is a (Hermitian) linear
combination of the mixed tensor products of the basis. Also in the
case of four qubit, it can be shown that the MMES (\ref{mmes4})
cannot be a non-degenerate ground state of a local Hamiltonian. In
fact, it can be verified that the three- and four-qubits couplings
present in (\ref{projmmes4}) and in the other projections
$\mathcal{M}_4^i$ are absent in $H_\mathrm{M}$. As a consequence,
the expectation value of the residual Hamiltonian on a basis state
must be equal to that of the other three states. By an argument
similar to that used in the case of three qubits, we find that, if
the state (\ref{mmes4}) is the ground state of a two-body (not
necessarily \emph{local}) Hamiltonian, it is at least four-fold
degenerate. The search for the most general local Hamiltonian
having $|M_4^1\rangle$ as an eigenstate has led to 24 independent
terms which leave the state $|M_4^1\rangle$ invariant, except for
an overall constant. They are graphically illustrated in Fig.\
\ref{hamilt4}. For the meaning of the symbols, see the caption.

The most evident characteristic of these terms is the abundance of
interactions coupling Pauli matrices along different axis. The
presence of these interactions is necessary if we want the
eigenstate $|M_4^1\rangle$ to be non-degenerate. General
conditions for non-degeneracy are very difficult to find if the
coupling parameters are generic. Such conditions can be obtained
more easily in a simplified Hamiltonian, which depends only on a
small number of parameters. For example, we have \barr
\label{Hjk4}
H_{Jk}&=&J(\sigma_4^x\sigma_1^z+\sigma_3^x\sigma_2^z)+k\bigl(\sigma_1^x\sigma_4^z+\sigma_2^x\sigma_3^z\nonumber\\
&+&\sigma_2^x\sigma_1^z+\sigma_1^x\sigma_2^z- \sum_{i=1}^4
\sigma_i^z \bigr) . \earr The Hamiltonian (\ref{Hjk4}) contains
interactions of the form $\sigma_i^x\sigma_j^z$ and the coupling
to an external field along the $z$-axis. We notice that it does
not couple qubits 3 and 4 to each other and, therefore, it can be
implemented on an open chain rather than on a ring. The eigenstate
$|M_4^1\rangle$, corresponding to the eigenvalue $2J$, is
non-degenerate if the following conditions hold
\begin{equation}
J\neq 0, \quad k\neq 0, \quad J\neq\pm\sqrt{\frac{3}{2}}k, \quad
J\neq\pm\sqrt{3}k.
\end{equation}
Despite the model dependence on a small number of parameters, it
is not easy to analytically determine the position of the excited
eigenstate $|M_4^1\rangle$ in the spectrum. Thus, we turned to a
numerical analysis: after generating $4\cdot 10^4$ random coupling
parameters (uniformly distributed in $[-1,1]^2$), we found that
the average position of the eigenvalue $2J$ is in the centre of
the energetic band; in the best case $|M_4^1\rangle$ is the second
excited state (as it cannot be the ground state).

\subsection{Five qubits}
For a system of five qubits, where perfect MMESs do exist \cite{MMES}, we
proceed in the same way as in the case of four qubits. We again
choose to consider a uniform MMES \cite{multent}
\begin{equation}\label{mmes5}
|M_5^1\rangle=\frac{1}{4\sqrt{2}}\sum_{i=0}^{31}\zeta^{(5)}_k
|k\rangle ,
\end{equation}
with
\begin{eqnarray}
\zeta^{(5)} &=&\{1, 1, 1, 1, 1, -1, -1, 1, 1, -1, -1, 1, 1, 1, 1,1, 1,1,\nonumber\\
& &-1, -1, 1, \ -1, 1, -1, -1, 1, -1, 1, -1, -1, 1, 1\}.\nonumber\\
\end{eqnarray}
The purity of the state $|M_5^1\rangle$ is minimal for any
bipartition and such a state is, therefore, a perfect MMES. As in
the case of four qubits, we can construct a basis of the Hilbert
space of the system which is formed only by MMES. The most general
Hamiltonian acting on the Hilbert space of the system is
\begin{equation}
H=\sum_{i=1}^{32} \epsilon_i \mathcal{M}_5^i +H_\mathrm{M} ,
\end{equation}
where $\mathcal{M}_5^i=|M_5^i\rangle\langle M_5^i|$ is a
projection and $H_\mathrm{M}$ is an Hermitian linear combination
of the mixed tensor products of the basis vectors.

In the case of five qubits the expressions are very complicated.
Notwithstanding this, we can apply the same procedure followed in
the previous subsection and draw some conclusions. Hamiltonians
containing only two-body (not necessarily local) interactions have
the same zero expectation value on each MMES of the chosen basis.
Thus, none of them can be the non-degenerate ground state of such
Hamiltonian.

\begin{figure}
\includegraphics[width=0.4\textwidth]{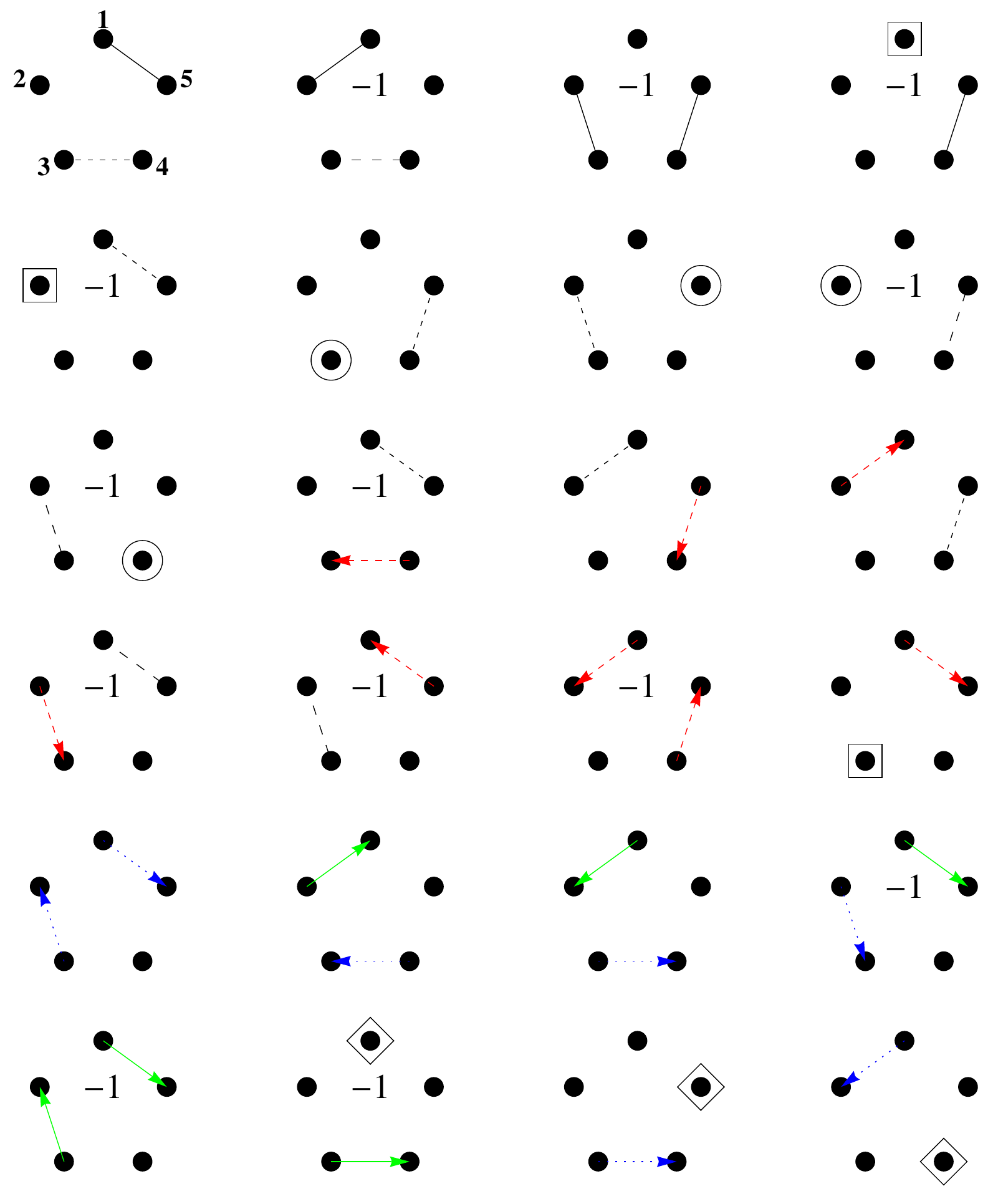}
\caption{(Color online) Graphic representation of all the
elementary local Hamiltonians having $|M_5^1\rangle$ as an
eigenstate. The meaning of the symbols is analogous to that in
Fig.\ \ref{hamilt4}.} \label{hamilt5}
\end{figure}

We found all the elementary Hamiltonians, containing only local
interactions, for which $|M_5^i\rangle$ is an eigenstate: they are
graphically represented in Fig. \ref{hamilt5}. It has been found
that, by combining the first nine terms with equal coupling
constants, the eigenstate $|M_5^i\rangle$ is non-degenerate. Thus,
we can obtain a non-degenerate MMES, eigenstate of a local
Hamiltonian, for a system of five qubits without considering mixed
interactions, as in the case of a four-qubit MMES. Moreover, we
observed that removing any one of these nine terms is incompatible
with the non-degeneracy of the MMES eigenstate. A numerical
analysis based on a random sampling of $10^5$ sets of coupling
parameters has confirmed (as in the four-qubits case) that the
MMES $|M_5^1\rangle$ is placed at the center of the energetic
band. It turns out to be impossible to reach one of the low-lying
excited states (we could not do better than placing the MMES at
the 14th excited level).

\section{Conclusions}

We have investigated whether it is possible to obtain $n$-qubit
MMES (for $2\leq n \leq 5$) as eigenstates of Hamiltonians
involving only local (few-body and nearest-neighbors) interactions
and fields. Since MMES exhibit very distributed non-local
correlations, the answer to this problem is nontrivial. We found
that MMES are the non-degenerate ground state only for $n=2$.
Already for $n=3$ the requirement that a MMES be the
non-degenerate ground state must be relaxed, and one finds that
MMES can at most be the first non-degenerate excited state. This
can be interpreted as a manifestation of entanglement frustration
\cite{nielsen1, nielsen2,NJP}. For $n=4$ we found, in a restricted
family of Hamiltonians, a MMES only as the second non-degenerate
excited eigenstate.

Besides its foundational interest, the present work is also
motivated by few-qubit applications. Most, if not all, practically
realizable quantum tasks involve only a very small number of
qubits. The most advanced quantum applications require that these
qubits be prepared in highly entangled states with high fidelity.
One expects that more performing applications would become
possible by making use of MMES: some examples were proposed in
\cite{Adesso,FFLMP,zhasong}. This clearly calls for efficient
methods to prepare/generate MMES with large yield and efficiency.
For example, if a MMES were the ground state of some Hamiltonian
$H$, then clearly one could engineer it by constructing $H$ and
letting the system relax toward its ground state. However, since
this is not the case, one must consider a partial relaxation
combined with control techniques or devise alternative strategies.
Another possible mechanism for generating MMES is dynamical rather
than static. This is obviously related to the degree of complexity
of a quantum circuit that generates MMES using, for instance,
two-qubit gates. Future investigations will be surely devoted to
this subject.

From the above mentioned point of view, due to the small number of
qubits considered, it would be interesting to study the
possibility of using presently realizable quantum systems with a
proper engineering of the interaction terms for generating MMES.
Finally, the physical conditions and strategies that would enable
one to efficiently prepare sates with large entanglement, such as
MMES, can be clearly generalized to other classes of entangled
states. Work is in progress in this direction.

\section*{ACKNOWLEDGEMENTS}
P.F. and G.F. acknowledge support through the project IDEA of
Universit\`a di Bari.

%%%%%%%%%%%%%%%%%%%%%%%%%%%%%%%%%%%%

\end{document}